\documentclass{PoS}

\title{A test of first order scaling in Nf =2 QCD:\\ a progress report }

\ShortTitle{A test of first order scaling in Nf =2 QCD: a progress report }

\author{\speaker{Guido Cossu}\\
        Scuola Normale Superiore, Dip. Fisica \& INFN, Pisa, Italy\\
        E-mail: \email{g.cossu@sns.it}}

\author{Claudio Bonati\\
        Dip. Fisica \& INFN, Pisa, Largo Pontecorvo 3, I-56127 Pisa, Italy\\
        E-mail: \email{bonati@df.unipi.it}}

\author{Massimo D'Elia\\
  Dip. Fisica \& INFN, Genova, Via Dodecaneso 33, I-16146 Genova, Italy\\
  E-mail: \email{delia@ge.infn.it}}

\author{Adriano Di Giacomo\\
  Dip. Fisica \& INFN, Pisa, Largo Pontecorvo 3, I-56127 Pisa, Italy\\
  E-mail: \email{digiaco@df.unipi.it}}

\author{Claudio Pica\\
  Brookhaven National Laboratory, Physics Department, Upton, NY 11973-5000, USA\\
  E-mail: \email{pica@bnl.gov}}

\abstract{We present the status of our analysis on the order of the finite temperature transition in QCD with two flavors of degenerate fermions. Our new simulations on large lattices support the hypothesis of the first order nature of the transition, showing a preliminary two state signal. We will discuss the implications and the next steps in our analysis.}

\FullConference{The XXVI International Symposium on Lattice Field Theory \\
July 14 - 19, 2008\\
Williamsburg, Virginia, USA}

\begin{document}

\section{Motivation}

It's a long standing issue whether the finite temperature transition in QCD (at zero chemical potential) from a confined to a deconfined phase is really a true transition or a simple crossover at finite fermion masses. This is not just an academic question if one wants to pursue the idea of a dual symmetry responsible for confinement, as suggested by experiments on the free quarks abundance in nature \cite{Amsler:2008zz}. In this respect, the most natural hypothesis  is the presence of a true phase transition even at finite masses rather than a crossover.

For three or more degenerate fermions we know that a first order transition occurs in a region nearby the chiral point \cite{Brown:1990ev,Liao:2001en}. In the 2+1 case there are some hints in favor of a crossover scenario, at physical quark masses, from the study of the susceptibility of the chiral condensate \cite{Aoki:2006we,Aoki:2006br}. The most debated case is QCD with two degenerate fermions: previous studies \cite{Aoki:1998wg, Karsch:1994hm,Bernard:1993en} -- performed on lattices with a modest extent -- claimed that the chiral transition belongs to the $O(4)$ universality class, thus leading to a crossover at finite quark masses. Such a conclusion implies also the presence of a possible second order end point in the temperature-chemical potential plane that should have a clear experimental signature (see for example \cite{Stephanov:2007fk, Stephanov:1998dy}). Despite the efforts, no evidence for such an end point has been found so far at heavy ions colliders (BNL-RHIC, CERN-SPS, GSI-FAIR). For all the above reasons, to establish the order of the chiral transition in two-flavor QCD is a particularly important problem in lattice QCD which is still open and deserves a careful and deep analysis. Starting from our previous investigation, we report here about the progress made during the last year, using larger lattices.

\section{Current status}

This work is the last step of a long-term project. The first step \cite{D'Elia:2005bv} was a direct check of a second order scaling. Pisarski and Wilczek \cite{Pisarski:1983ms} predicted, by means of a chiral model (thus assuming that only the chiral degrees of freedom are responsible for the order of the transition), that the QCD transition at the chiral point should be in the $O(4)$ universality class if a IR fixed point exists\footnote{In the case of staggered fermions this universality class reduces to $O(2)$ at finite lattice spacing.}, i.e. if it is a second order transition, not excluding the first order case.

\begin{figure}[t]
  \begin{centering}
    \includegraphics[clip=true,width=.48\textwidth]{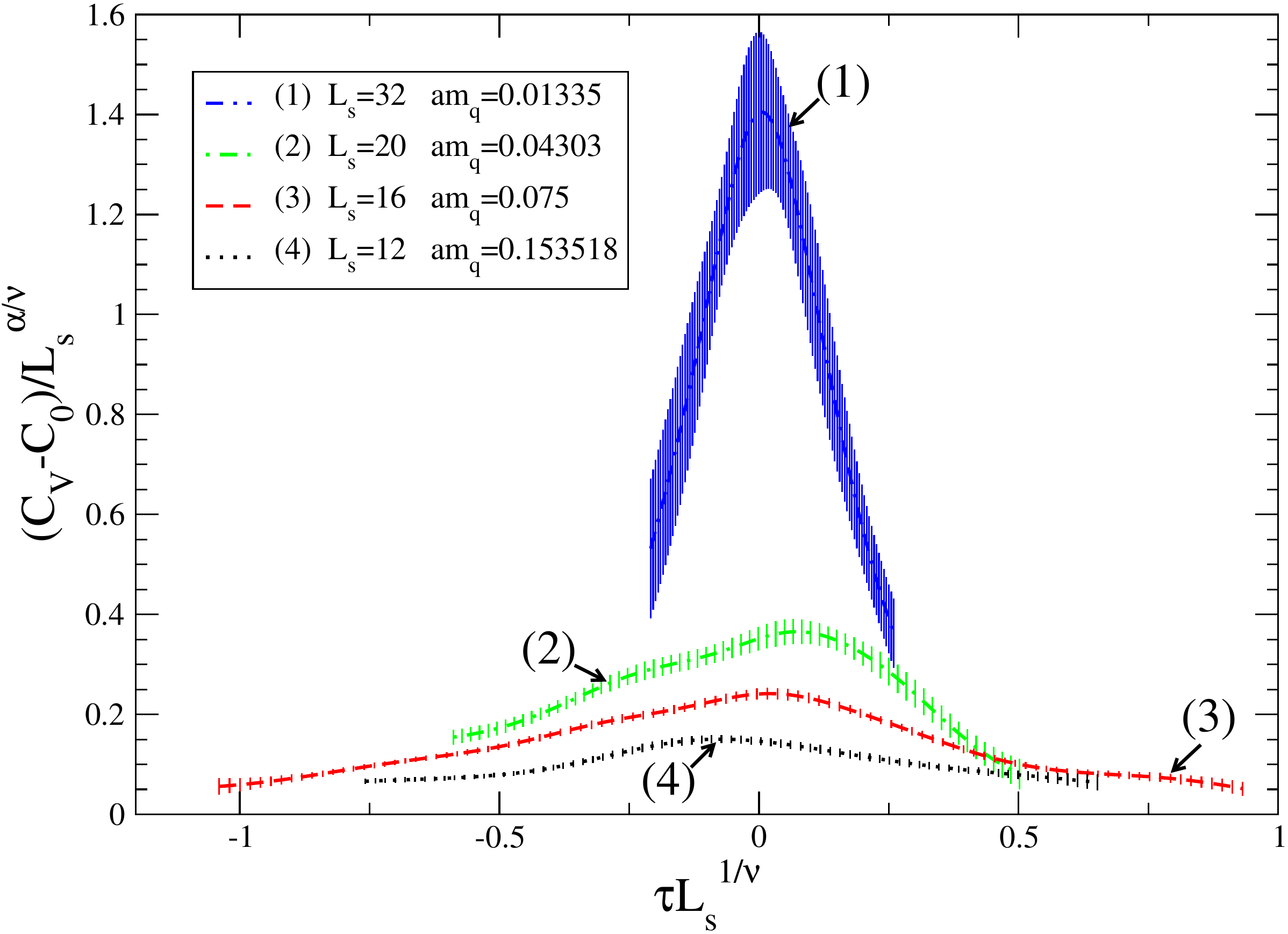}
    \includegraphics[clip=true,width=.48\textwidth]{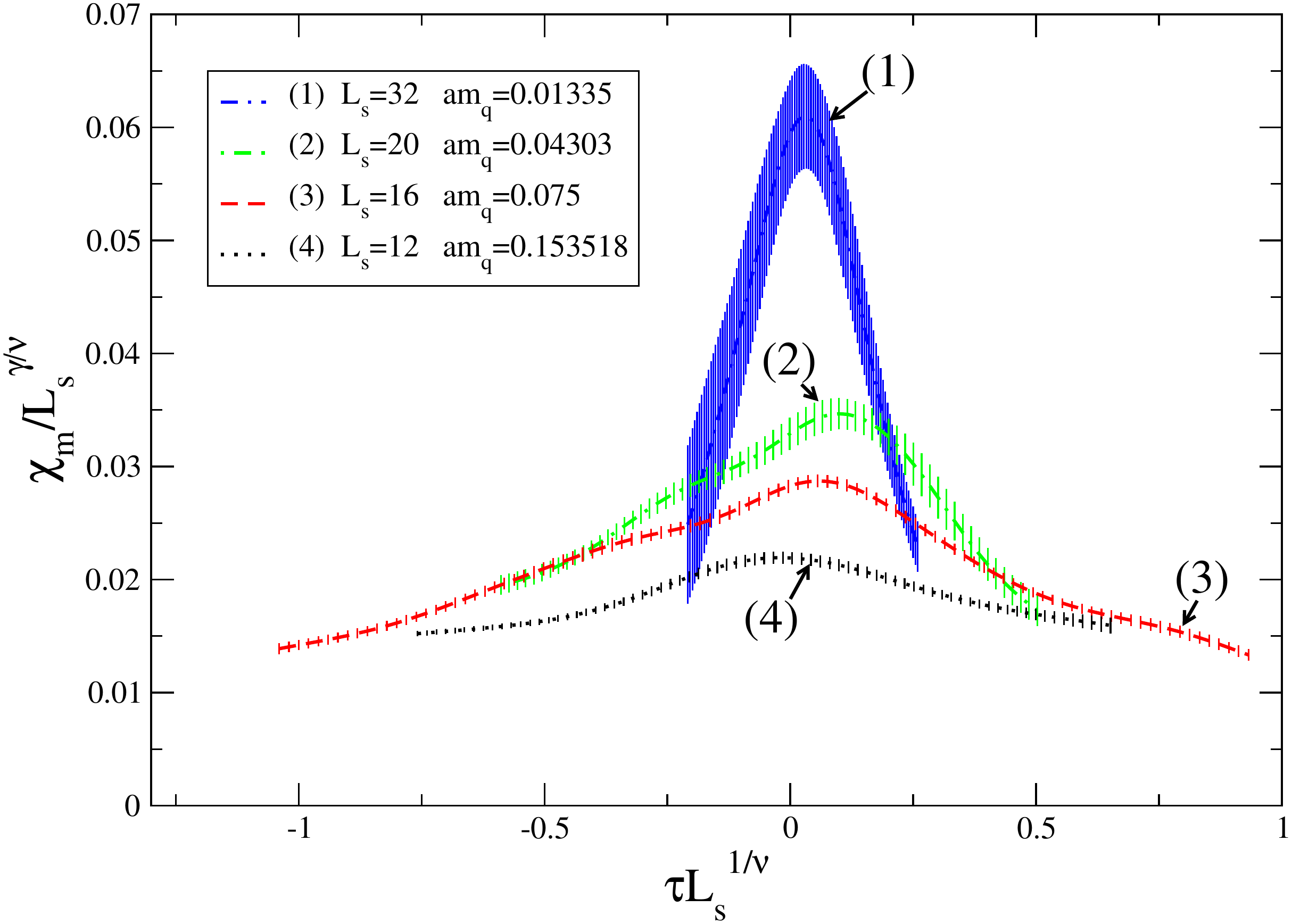}\\
    \vspace{5pt}
    \includegraphics[clip=true,width=.48\textwidth]{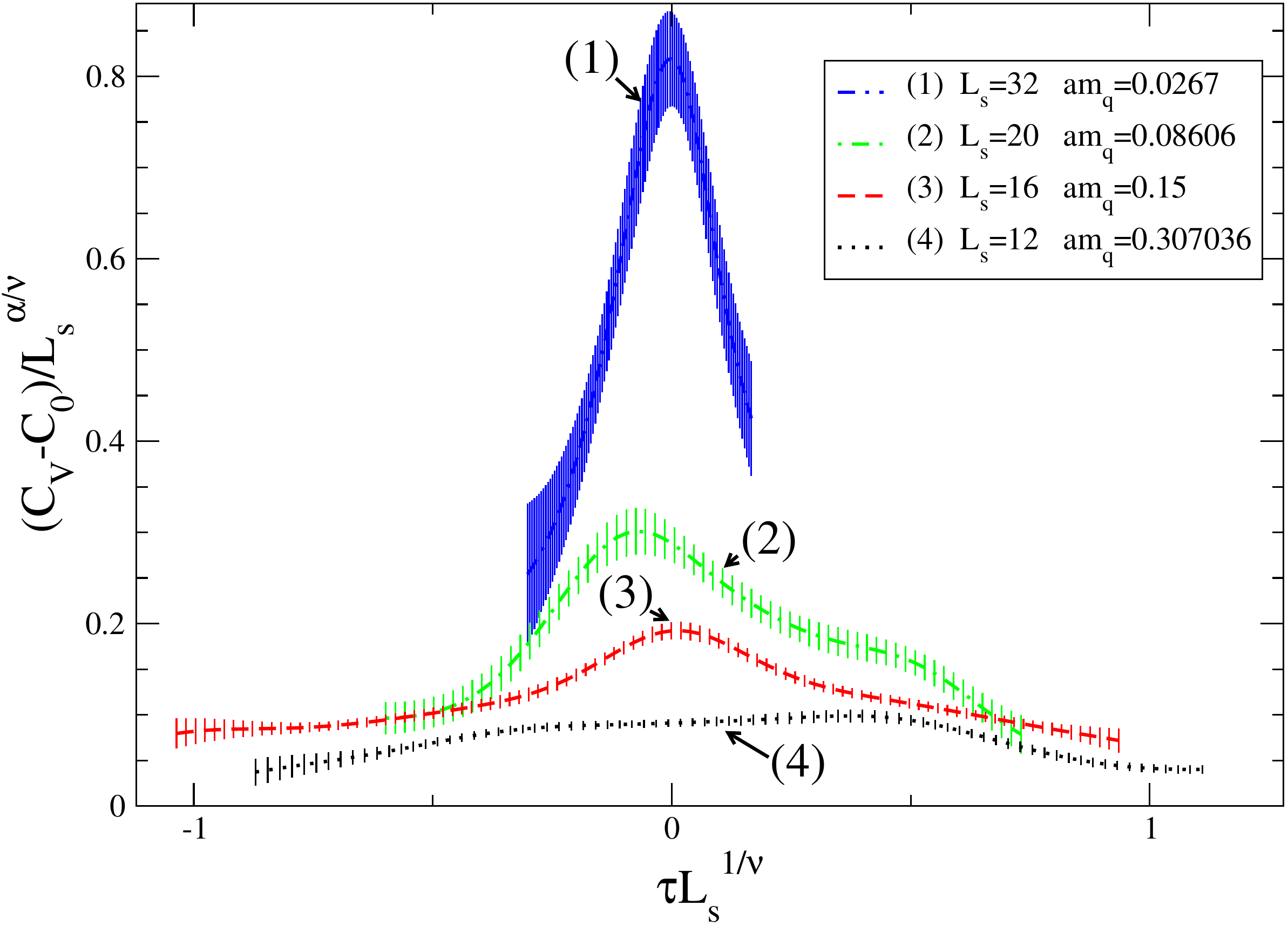}
    \includegraphics[clip=true,width=.48\textwidth]{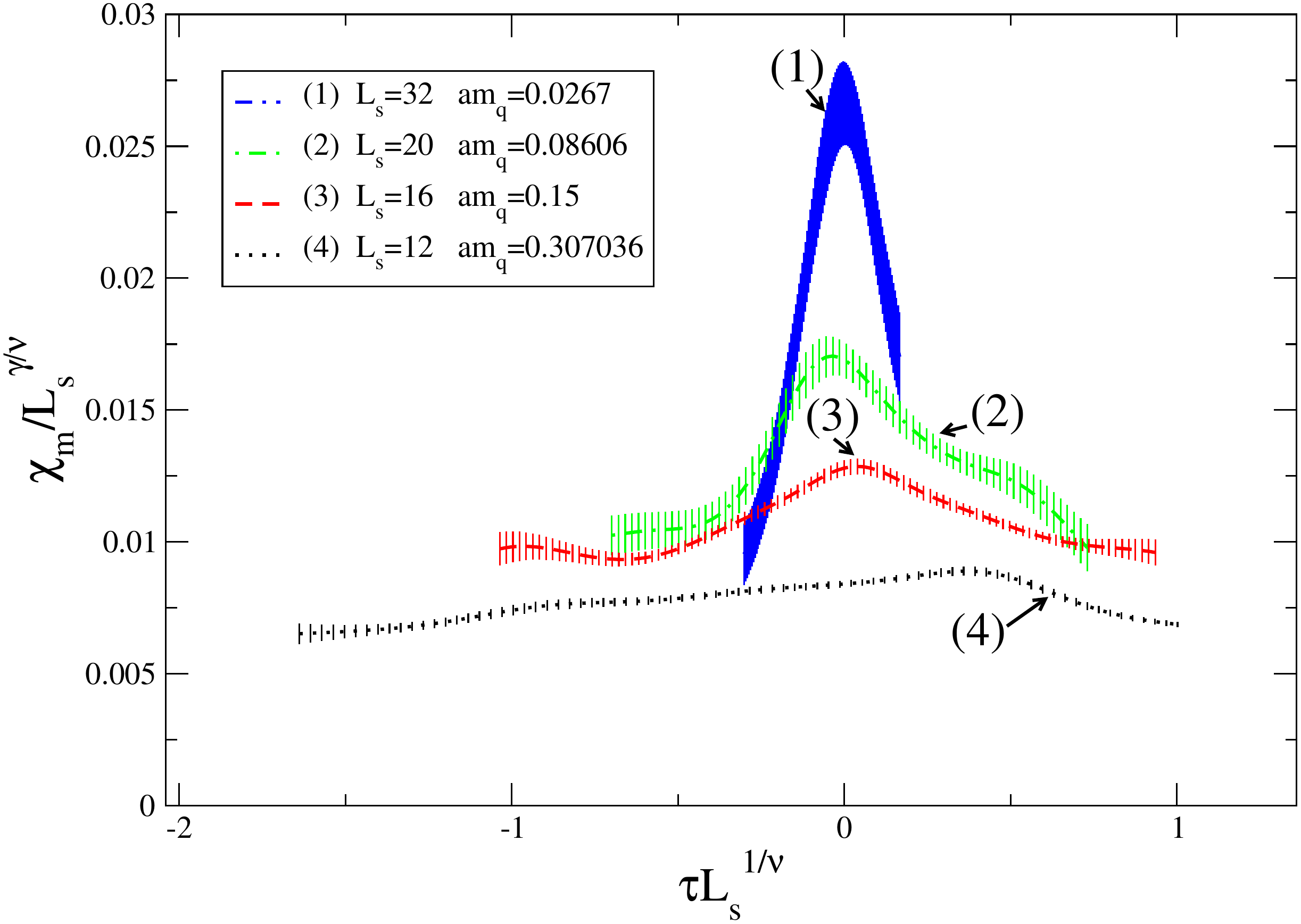}\\
    \caption{Second order consistency check. Upper row refers to $O(4)$ scaling, lower to $O(2)$. Columns refer respectively to the specific heat and the chiral susceptibility from left to right.}
    \label{Secondorder}
  \end{centering}
\end{figure}

A finite size scaling (FSS) analysis is the best way to address the problem of determining the critical exponents of a transition in a lattice simulation. The problem has two scales, the temperature and the bare quark mass, as shown e.g. in the equations of the scaling of the free energy density and its derivative, the specific heat:
\begin{eqnarray}
L/kT &\simeq & L_s^d \phi(\tau L_s^{1/\nu},am_qL_s^{y_h})\\
C_v -C_0 &\simeq & L_s^{\alpha/\nu} \phi_c(\tau L_s^{1/\nu},am_qL_s^{y_h})\label{SH}
\end{eqnarray}
Our strategy is to get rid of one of the two scales, namely the second one, by fixing its value $am_qL_s^{y_h} = \rm{const}$, and looking at the FSS in the other variable. Obviously this strategy implies previous knowledge of $y_h$ so one could test consistency of data with a particular scaling hypothesis. The result of these consistency tests are shown in figure \ref{Secondorder}. If the $O(4)$, or $O(2)$, scaling is correct then all the curves should fall on top of each other. The conclusion based on these first observations is that a second order universality class as predicted by chiral models is excluded. Notice also that the universality classes considered have $\alpha < 0$, i.e. the specific heat should not grow with the volume, in contrast with data.
Notice that O(4) and O(2) critical indexes are both very close to those of the $U(2)_L \times U(2)_R/U(2)_V$ universality class, predicted as a possible alternative in case of a light $\eta'$ meson at the transition~\cite{Basile:2005hw}. Therefore our data exclude that universality class as well.

\begin{figure}[h]
  \begin{centering}
    \includegraphics[clip=true,width=.48\textwidth]{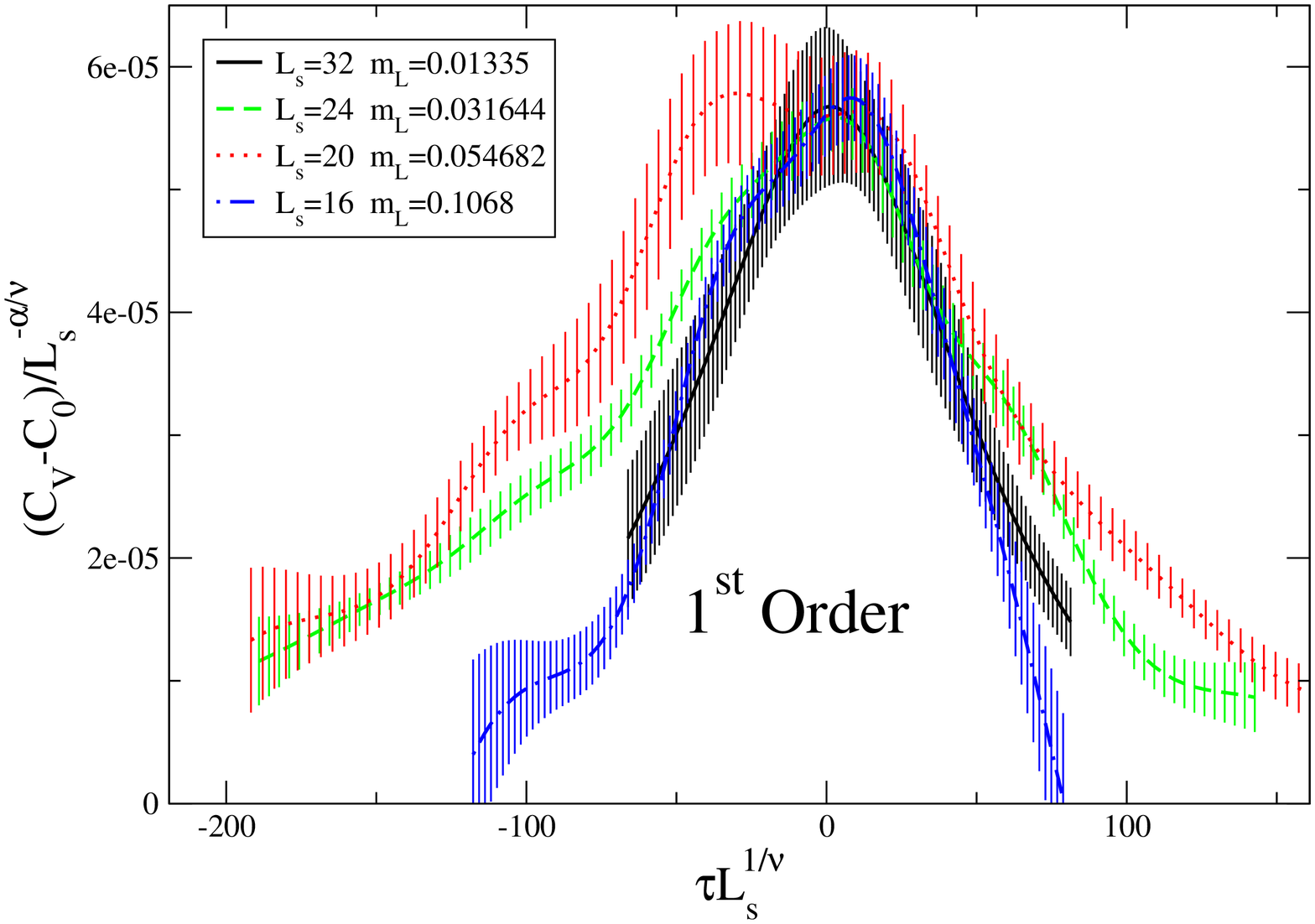}
    \includegraphics[clip=true,width=.48\textwidth]{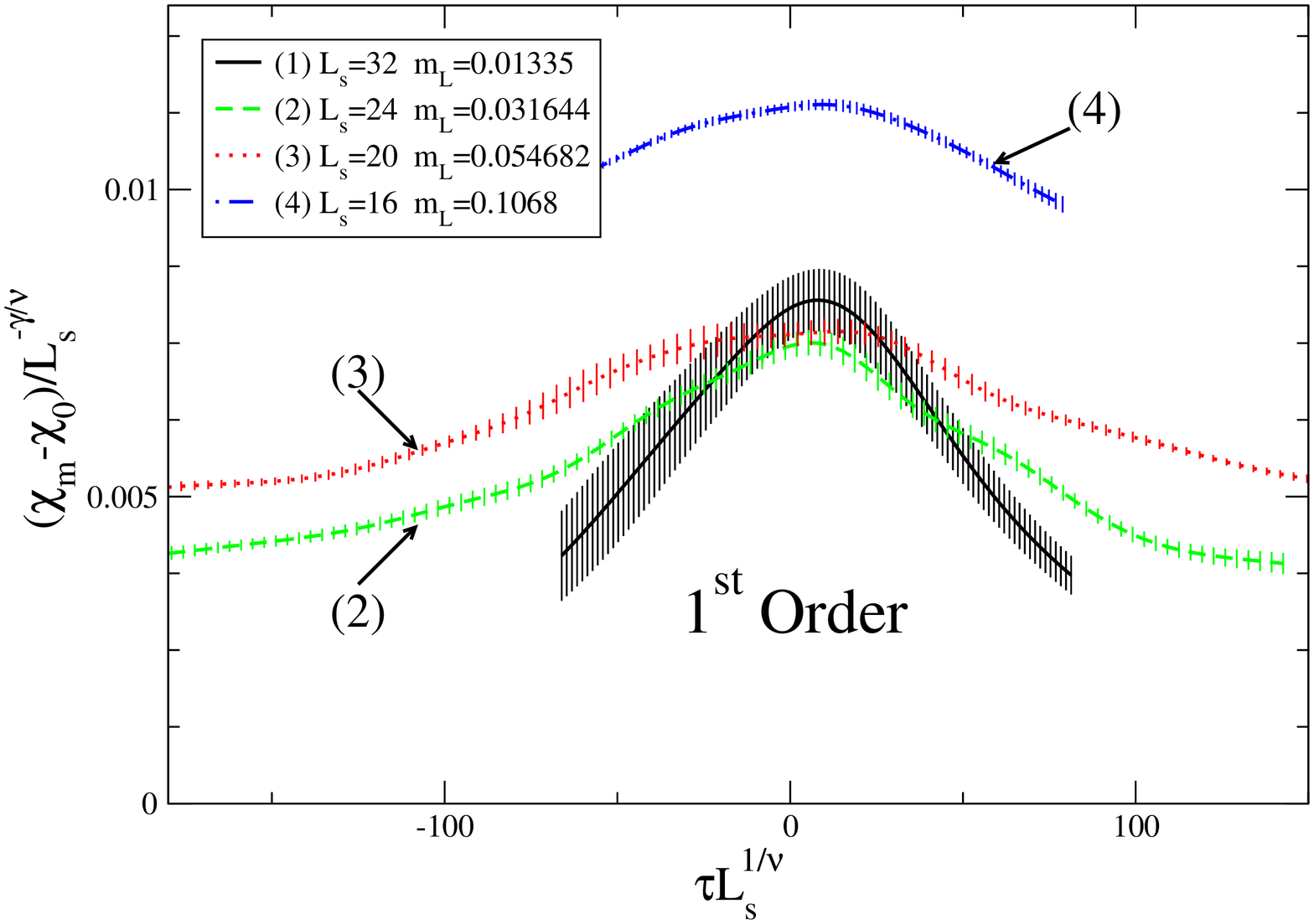}
    \caption{First order consistency check of the specific heat (left) and the chiral susceptibility (right).}
    \label{Firstorder}
  \end{centering}
\end{figure}

The following step is to check directly the first order scaling for which some hints are found using approximate scaling laws in \cite{D'Elia:2005bv}. The results of such a test \cite{Cossu:2007mn},  where $y_h=3$, are shown in figure \ref{Firstorder}. The specific heat scales quite nicely with the first order hypothesis and also the chiral condensate susceptibility is in agreement if one excludes the curve at $L_s=16$ arguing that it probably lies outside the scaling region since it has the smallest volume and the heaviest mass.

\section{First order scaling analysis}
The preceding observations were convincing enough to argue in favor of a first order transition. However several questions still remain:
\begin{itemize}
\item where is the scaling with the volume of the peak at fixed mass, expected for some small masses? (here is difficult to quantify the term ``small'')
\item where are the double peaks expected in the observables histograms near the transition point?
\end{itemize}
Before looking at the simulations let's consider again the scaling laws at fixed bare quark mass and in particular equation \ref{SH}. If a second order transition is present at the chiral point then at finite mass everything is analytical and no divergence can arise, i.e. in the thermodynamical limit any dependence on $L$ should vanish. We can expand in terms of the inverse of the second parameter and find that the leading term in the expansion of $\phi_c$ must be $\propto 1/(am_q L^{y_h})^{\alpha/(\nu y_h)}$. In the case of a first order transition, where the equations are valid if the transition is really weak (as this is the case), a constant term (in volume) is present because of a peculiar cancellation occurring only in this case:
\begin{equation}
C_V-C_0 \simeq am_q^{-1}\phi_1(\tau V) + V \phi_2(\tau V) 
\end{equation}
the second term giving non-zero latent heat.

The relative weight of the two terms is unknown a priori. It is perfectly possible that the singular, diverging, term is really small at the volumes explored. Nevertheless one should observe a shrinking with the volume in the width of the specific heat curve, in contrast with the crossover case where a constant curve is expected.

In order to check volume effects we decided to dedicate a large part of our computational facilities to a run with standard staggered fermions at $am_q=0.01335$ on a $48^3\times 4$ lattice. This corresponds to a pion mass of about twice the physical value and to a spatial size of $\sim 13-14$ fm. We present here our results at this stage. 

\begin{figure}[h]
  \begin{centering}
    \includegraphics[clip=true,width=.48\textwidth]{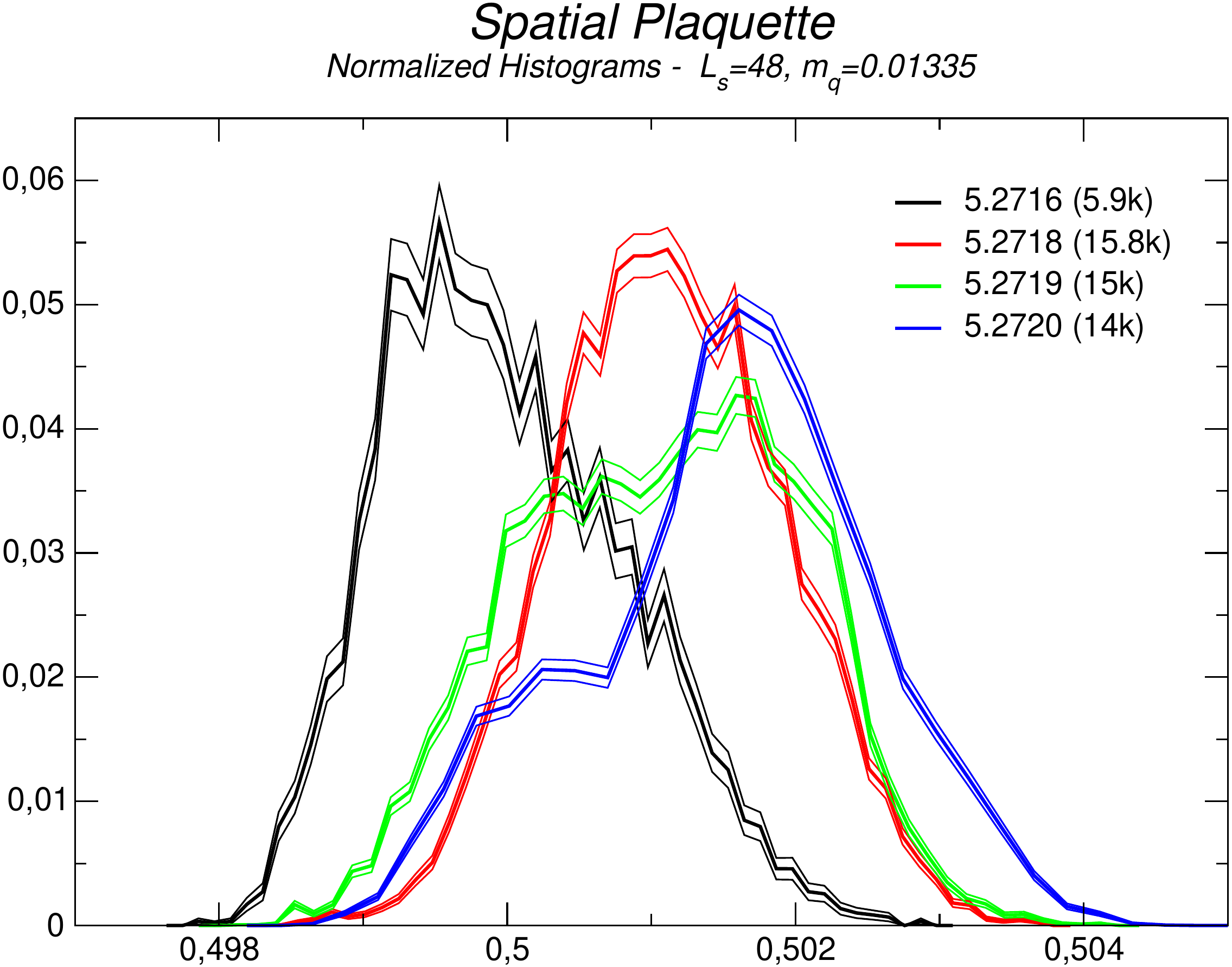}
    \includegraphics[clip=true,width=.48\textwidth]{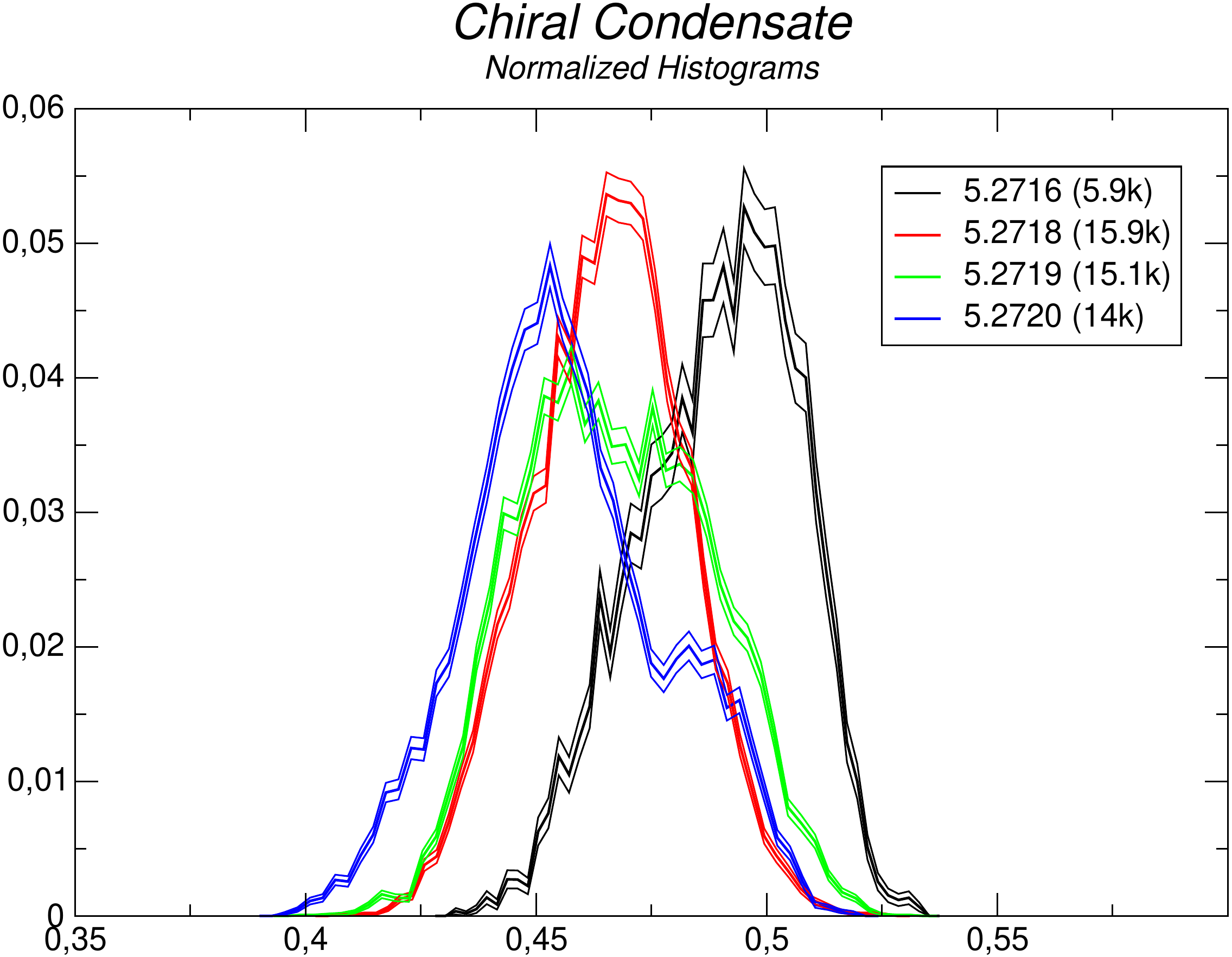}
    \caption{Spatial plaquette (left) and chiral condensate (right) distribution histograms.}
    \label{Histograms}
  \end{centering}
\end{figure}

\begin{figure}[h]
  \begin{centering}
    \includegraphics[clip=true,width=.48\textwidth]{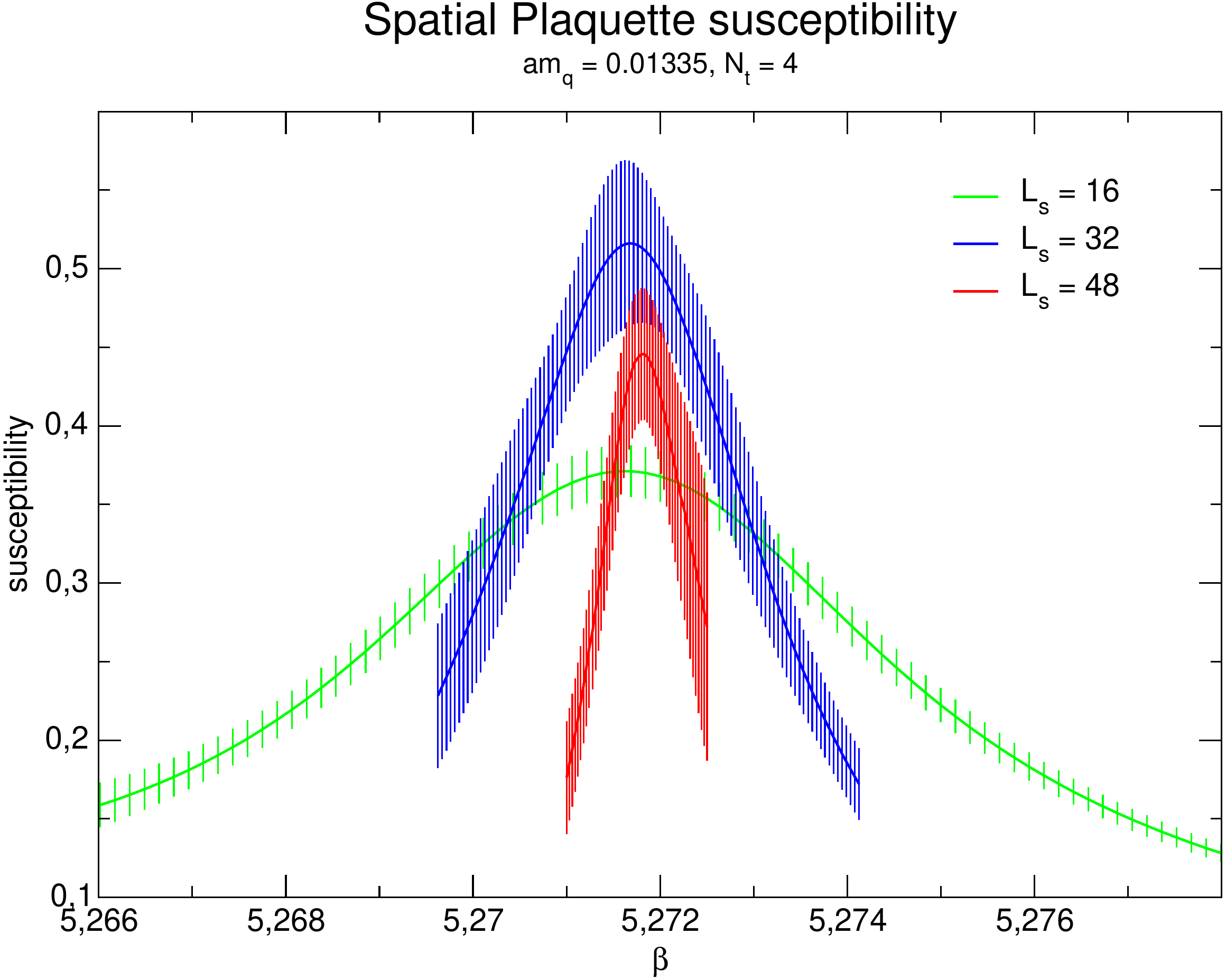}
    \includegraphics[clip=true,width=.48\textwidth]{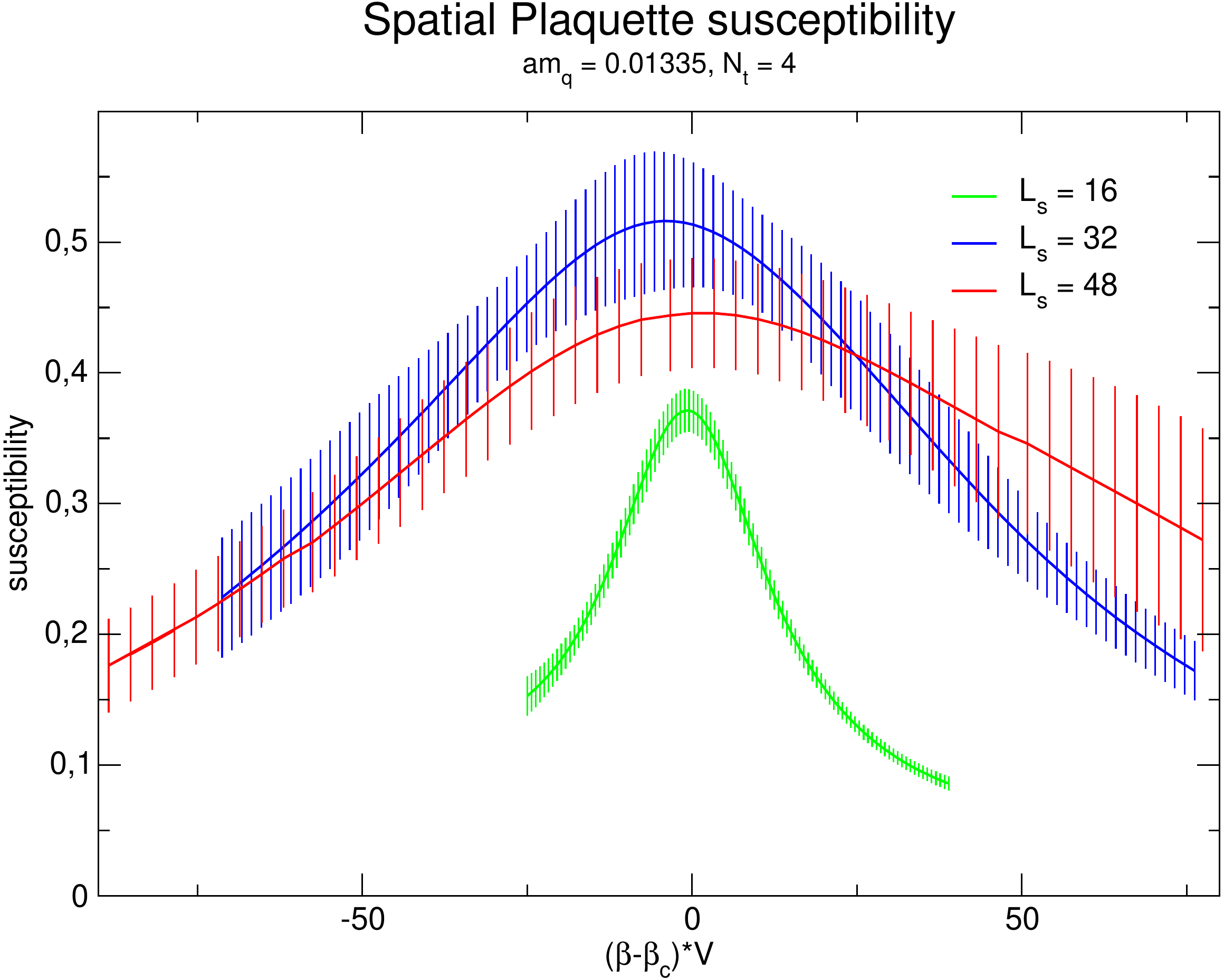}
    \caption{Spatial plaquette susceptibility (left). Rescaling of $\beta$ axis (right) showing a good scaling of peak width. The $L_s=16$ lattice is out of the scaling region.}
    \label{Susc}
  \end{centering}
\end{figure}

We simulated four different temperatures around the peak of the specific heat. In figure \ref{Histograms} we show the plaquette and the chiral condensate histograms.
Data seem to present a weak signal of a double peak for betas 5.2719 and 5.2720. However statistics is still too low to draw any conclusion in this respect. A close look at the spatial plaquette susceptibility ($\sim$ specific heat) (fig. \ref{Susc}) shows that the peak does not grow with the volume but the reweighted curve shrinks with the correct factor (see right figure).
\section{Conclusions}
The determination of the order of the $N_f=2$ QCD transition is an interesting, rather tough problem, and we are still far from a conclusive answer.
We reported our progress in understanding various aspects of the problem. First we concluded that a second order phase transition in the $O(4)$ or $O(2)$ universality classes at the chiral point has to be excluded. Secondly, investigating the possibility of a first order nature of the transition, we found several indications that could be the right answer. A direct consistency check of this hypothesis gave a positive answer, especially for the specific heat, an observable that does not imply any assumption on the symmetry behind the transition. Looking for metastabilities proved to be a more difficult task. Very weak signals of double peaks were found. If the transition is really first order this implies a very small latent heat, so that double peak structures are masked by simple thermal noise. A simple analysis of the scaling equations shows that in the first order case we should expect two contributions to scaling, one regular and one singular giving the latent heat. The actual data sets suggest that the regular term is much bigger than the singular one at the volumes and masses explored. We observe the expected shrinking of the specific heat curve with the exponents predicted for a first order.

We need to increase statistics to have reliable ensemble with small errors and if our early conclusions are confirmed (eventually by simulations with an improved action and/or finer lattice spacing) the standard crossover scenario has to be changed.

This work was done using the apeNEXT facilities in Rome during the time of several months.


\begin{thebibliography}{99}
  %\cite{Amsler:2008zz}
\bibitem{Amsler:2008zz}
  C.~Amsler {\it et al.}  [Particle Data Group],
  %``Review of particle physics,''
  Phys.\ Lett.\  B {\bf 667}, 1 (2008).
  %%CITATION = PHLTA,B667,1;%%
  
  %\cite{Karsch:2001nf}
  % \bibitem{Karsch:2001nf}
  %   F.~Karsch, E.~Laermann and C.~Schmidt,
  %   ``The chiral critical point in 3-flavor QCD,''
  %   Phys.\ Lett.\  B {\bf 520}, 41 (2001)
  %   [arXiv:hep-lat/0107020].

\bibitem{Brown:1990ev}
   F.~R.~Brown {\it et al.},
   %``On the existence of a phase transition for QCD with three light quarks,'' 
   Phys.\ Rev.\ Lett.\ {\bf 65}, 2491 (1990).

  % \cite{Liao:2001en}
\bibitem{Liao:2001en}
  X.~Liao,
  % ``Study of 3-flavor QCD finite temperature phase transition with  staggered
  % fermions,''
  Nucl.\ Phys.\ Proc.\ Suppl.\  {\bf 106}, 426 (2002)
  [arXiv:hep-lat/0111013].
  %% CITATION = NUPHZ,106,426;%%
  %% CITATION = PHLTA,B520,41;%%
  
  % \cite{Aoki:2006we}
\bibitem{Aoki:2006we}
  Y.~Aoki, G.~Endrodi, Z.~Fodor, S.~D.~Katz and K.~K.~Szabo,
  % ``The order of the quantum chromodynamics transition predicted by the
  % standard model of particle physics,''
  Nature {\bf 443}, 675 (2006)
  [arXiv:hep-lat/0611014].
  %% CITATION = NATUA,443,675;%%
  
  % \cite{Aoki:2006br}
\bibitem{Aoki:2006br}
  Y.~Aoki, Z.~Fodor, S.~D.~Katz and K.~K.~Szabo,
  % ``The QCD transition temperature: Results with physical masses in the
  % continuum limit,''
  Phys.\ Lett.\  B {\bf 643}, 46 (2006)
  [arXiv:hep-lat/0609068].
  %% CITATION = PHLTA,B643,46;%%

  %\cite{Aoki:1998wg}
\bibitem{Aoki:1998wg}
  S.~Aoki {\it et al.}  [JLQCD Collaboration],
  %``Scaling study of the two-flavor chiral phase transition with the
  %Kogut-Susskind quark action in lattice QCD,''
  Phys.\ Rev.\  D {\bf 57}, 3910 (1998)
  [arXiv:hep-lat/9710048].
  %%CITATION = PHRVA,D57,3910;%%

  %\cite{Karsch:1994hm}
\bibitem{Karsch:1994hm}
  F.~Karsch and E.~Laermann,
  %``Susceptibilities, the specific heat and a cumulant in two flavor QCD,''
  Phys.\ Rev.\  D {\bf 50}, 6954 (1994)
  [arXiv:hep-lat/9406008].
  %%CITATION = PHRVA,D50,6954;%%

%\cite{Bernard:1993en}
\bibitem{Bernard:1993en}
  C.~W.~Bernard {\it et al.},
  %``The Nature of the thermal phase transition with Wilson quarks,''
  Phys.\ Rev.\  D {\bf 49}, 3574 (1994)
  [arXiv:hep-lat/9310023].
  %%CITATION = PHRVA,D49,3574;%%

  % \cite{Stephanov:2007fk}
\bibitem{Stephanov:2007fk}
  M.~A.~Stephanov,
  % ``QCD phase diagram: An overview,''
  PoS {\bf LAT2006}, 024 (2006)
  [arXiv:hep-lat/0701002].
  %% CITATION = POSCI,LAT2006,024;%%

  %\cite{Stephanov:1998dy}
\bibitem{Stephanov:1998dy}
  M.~A.~Stephanov, K.~Rajagopal and E.~V.~Shuryak,
  %``Signatures of the tricritical point in {QCD},''
  Phys.\ Rev.\ Lett.\  {\bf 81}, 4816 (1998)
  [arXiv:hep-ph/9806219].
  %%CITATION = PRLTA,81,4816;%%
  
  % \cite{D'Elia:2005bv}
\bibitem{D'Elia:2005bv}
  M.~D'Elia, A.~Di Giacomo and C.~Pica,
  % ``Two flavor QCD and confinement,''
  Phys.\ Rev.\  D {\bf 72}, 114510 (2005)
  [arXiv:hep-lat/0503030].
  %% CITATION = PHRVA,D72,114510;%%
  %\cite{Pisarski:1983ms}
\bibitem{Pisarski:1983ms}
  R.~D.~Pisarski and F.~Wilczek,
  %``Remarks On The Chiral Phase Transition In Chromodynamics,''
  Phys.\ Rev.\  D {\bf 29}, 338 (1984).
  %%CITATION = PHRVA,D29,338;%%

\bibitem{Basile:2005hw}
  F.~Basile, A.~Pelissetto and E.~Vicari,
  PoS {\bf LAT2005}, 199 (2006)
  [arXiv:hep-lat/0509018].

  %\cite{Cossu:2007mn}
\bibitem{Cossu:2007mn}
  G.~Cossu, M.~D'Elia, A.~Di Giacomo and C.~Pica,
  %``Two flavor QCD and confinement - II,''
  arXiv:0706.4470 [hep-lat].
  %%CITATION = ARXIV:0706.4470;%%



\end{thebibliography}
\end{document}